\documentclass[prb,aps,reprint]{revtex4-1}
\usepackage{graphicx}
\usepackage{latexsym}
\usepackage{amsmath}
\usepackage{subfigure}
\usepackage{amssymb}
\usepackage{amsfonts}
\usepackage{mathrsfs}
\usepackage{dcolumn}
\usepackage{bm}

\newcommand\vccsi[1]{(\text{V}$_\text{C}$C$_\text{Si}$)$^{#1}$}
\newcommand\vsi[1]{\text{V}$_\text{Si}^{#1}$}
\newcommand\csplit[1]{\text{C}$_\text{split}^{#1}$}

\begin{document}

\title{Energetics and metastability of the silicon vacancy in cubic SiC}

\author{Fabien Bruneval}
\author{Guido Roma}

\affiliation{CEA, DEN, Service de Recherches de M\'etallurgie Physique, F-91191 Gif-sur-Yvette, France}

\date{\today}

\begin{abstract}
The silicon vacancy is a prominent intrinsic defect of cubic SiC (3C-SiC) to which much effort has been devoted so far, experimentally and theoretically.
We calculate its properties using the $GW$ approximation that does not suffer from the band gap problem.
The obtained formation and transition energies deviate significantly from the usual density functional theory evaluations
and now compare favorably with experiment.
A new assignment for the main line of photoluminescence is then proposed.
We further perform $GW$ calculations for the saddle point of reaction paths.
The resulting barrier energies explain the thermal annealing experiments thanks to an original mechanism mediated by
a minority charge configuration.
\end{abstract}

\maketitle

\section{Introduction}

Cubic silicon carbide (3C-SiC) is considered as a promising semiconductor for use in severe environments.
The potential applications include hardened electronic devices and nuclear fuel coatings.
In particular, its temperature and radiation resistance makes 3C-SiC attractive
for the next generation of high-temperature gas-cooled nuclear reactors \cite{yvon_jnm2009}.
Due to these purposes, irradiation-induced point defects of 3C-SiC have been widely studied both
experimentally  \cite{itoh_jap1989,itoh_jem1992,itoh_jap1995} and theoretically \cite{torpo_apl1999,petrenko_jpcm2002,bockstedte_prb2003}.
Among the intrinsic defects, the silicon vacancy plays a prominent role since it remains stable up to relatively high temperatures
and it is easily identified by electron paramagnetic resonance (EPR) measurements due to its high-spin configuration \cite{itoh_jap1989}.
Unfortunately, a careful comparison between the experimental data and the \textit{ab initio} calculations obtained within density functional theory (DFT)
is not satisfactory.

The usual approximations of DFT suffer from the infamous band gap problem.
There is nowadays a common agreement that the band gap problem particularly plagues DFT predictions
of defect properties in semiconductors and insulators \cite{lambrecht_pssb2011}.
Point defects in non metallic solids generally produce additional electronic levels inside the band gap region.
The precise location of these levels is critical for most defect properties:
charge state transition energy, formation energy, photoluminescence (PL) lines.
Predictive calculations absolutely need to be based on schemes that are devoid of the band gap problem.
The $GW$ approximation of the many-body perturbation theory \cite{hedin_pr1965} has been proven
for several years to yield the correct band gaps \cite{hybertsen_prl1985,aulbur_2000}.
Still the $GW$ calculation of defects is a great numerical challenge.

In the present article, we calculate, using the $GW$ approximation, the energetics and the stability of the silicon vacancy in 3C-SiC.
In order to access these properties, we performed the $GW$ calculations not only for equilibrium structures, but also
for saddle point positions.
The $GW$ results were combined so to reduce an error, named concavity, that was recently identified for the $GW$ approximation
by one of us \cite{bruneval_prl2009}.
The procedure, explained in Sec.~\ref{sec:method}, allows us to calculate the formation energies of the metastable form \vsi{}
and of the stable configuration \vccsi{} within the $GW$ framework in Sec.~\ref{sec:formation}.
The results bring a quantitative agreement with various experimental measurements:
EPR, PL, and thermal annealing.
Noticeably, the calculations propose a reassignment of the main PL line in Sec.~\ref{sec:pl}
and highlight an original decay mechanism through a minority charge state in Sec.~\ref{sec:annealing}.

\section{Method}
\label{sec:method}
\subsection{$GW$ approximation applied to the defects}

Thanks to its accuracy in predicting band gaps and electronic levels, the $GW$ approximation appears to be the method of choice to deal with point defects in semiconductors and insulators.
However some noticeable bottlenecks have inhibited the use of this scheme so far.
Very recent works \cite{rinke_prl2009,giantomassi_pssb2011,bruneval_prl2009,bruneval_prb2008}
have lifted the theoretical and numerical problems as recapitulated in the following.

Deriving from Green's function theory, the $GW$ approximation provides meaningful quasiparticle electronic levels.
The $GW$ approximation is hence reliable in predicting charge changes with constant geometry.
For instance, the energy of the highest occupied molecular orbital (HOMO) can be interpreted as a total energy difference at constant geometry,
which in turn defines the ionization potential $I$:
\begin{eqnarray}
 \epsilon_\text{HOMO}^{GW}(\text{V}_\text{Si}^{q},q)
     & = & E_0(\text{V}_\text{Si}^{q},q) - E_0(\text{V}_\text{Si}^{q},q+1)  \\
     & = & - I(\text{V}_\text{Si}^{q},q)   ,
\end{eqnarray}
where the first argument refers to the geometry of the defect and the second shows the actual charge state $q$.
$E_0$ stands for the ground-state total energy and $\epsilon$ for the quasiparticle energy.
Symmetrically, the lowest unoccupied molecular orbital (LUMO) energy is a total energy difference, which can be understood as the opposite of the
electron affinity: $-A$.
In the context of defects, these quasiparticle energies
are named vertical transition energies $\epsilon_v(q/q+1)$.
Note that this vertical transition energy $\epsilon_v(q/q+1)$ could be obtained in principle either from 
$\epsilon_\text{HOMO}^{GW}(\text{V}_\text{Si}^{q},q)$ or from  $\epsilon_\text{LUMO}^{GW}(\text{V}_\text{Si}^{q},q+1)$.
One of us showed recently \cite{bruneval_prl2009} that these two values slightly differ in practice due to the concavity
of the $GW$ approximation.
The most consistent way to handle this discrepancy is hence to approximate
the vertical transition energy by the mean value of the two mentioned quasiparticle energies.

The calculation of defect formation energies also requires to relax the geometries, since
the structure of a defect changes according to the charge state.
Unfortunately the $GW$ total energies and forces are out of reach even in state-of-the-art implementations.
That is why Rinke and coworkers \cite{rinke_prl2009} recently introduced a scheme that combines DFT and $GW$.
DFT is used to deal with the structural changes at constant charge state. The $GW$ approximation is employed for charge changes at constant geometry.
When traveling on a given Born-Oppenheimer surface, the band gap underestimation is expected to be harmless and therefore DFT can be safely used.
Then the $GW$ approximation allows one to address the vertical transitions from one Born-Oppenheimer surface to another one.
The combination of the two moves finally permits one to change both the geometry and the charge states.
The thermodynamic charge transition energies $\epsilon_{th}(q/q+1)$ can thus be obtained by introducing the energy of an intermediate point:
\begin{eqnarray}
 \epsilon_{th}(q/q+1) 
                   &=& E_0(\text{V}_\text{Si}^{q},q) - E_0(\text{V}_\text{Si}^{q+1},q)  \nonumber \\
                   & & + E_0(\text{V}_\text{Si}^{q+1},q) - E_0(\text{V}_\text{Si}^{q+1},q+1) . \label{eq:epsilon_a} 
\end{eqnarray}
The first two terms account for a structural change at constant charge $q$, which is obtained accurately from two DFT calculations.
The last two terms account for a charge change at constant structure, which is calculated from $GW$ quasiparticle energies as explained
above.

With the present combined DFT/$GW$ approach, the absolute formation energy of the defect for any charge state can be obtained,
provided one reference formation energy. In Ref.~\onlinecite{rinke_prl2009} the reference formation energy was chosen to be the
DFT total energy with all defect states in the band gap empty, so that the band gap problem does not enter in the total energy.
Then the neighboring charge state formation energy can be obtained as exemplified here:
\begin{multline}
 E_f(\text{V}_\text{Si}^{+}) = E_f(\text{V}_\text{Si}^{2+}) 
              + \epsilon_{th}(1+/2+) 
                    - \epsilon_\text{VBM}^{GW} - \mu_e ,
\end{multline}
where $E_f$ stands for the usual formation energy
and $\mu_e$ stands for the Fermi level with its zero set at the $GW$ valence band maximum, $\epsilon_\text{VBM}^{GW}$.

\subsection{Technical details}

The main numerical bottleneck of the $GW$ approximation comes from
the dependence of the $GW$ exchange-correlation self-energy onto the empty states.
This problem was given a partial answer in Ref.~\onlinecite{bruneval_prb2008}, which allows
us to run $GW$ calculations for supercells as large as 215 atoms using a very low 3:1 ratio between empty and occupied states.

We employ the usual perturbative $GW$ method ($G_0W_0$) \cite{hybertsen_prl1985} with norm-conserving pseudopotentials and the Godby-Needs plasmon-pole model,
using a 30~Ha cutoff for the wavefunctions and a 6~Ha cutoff for the dielectric matrix.
The k-point grid is a 2x2x2 Monkhorst-Pack grid for the DFT force calculations. It is reduced to the $\Gamma$ point only when turning to $GW$ calculations.
The calculations are spin polarized and the neutral vacancy is approximated by its triplet spin state, which may overestimate its formation
energy by less than 0.1~eV.\cite{deak_apl1999,zywietz_prb2000}
The obtained band gap is 1.35~eV within the local density approximation (LDA), 2.19~eV within $GW$ to be compared to the experimental value, 2.37~eV.
The discrepancy between the $GW$ band gap and the experimental band gap can be considered as the error bar of the results presented in the following.

\begin{figure}[t]
\includegraphics[width=0.95\columnwidth]{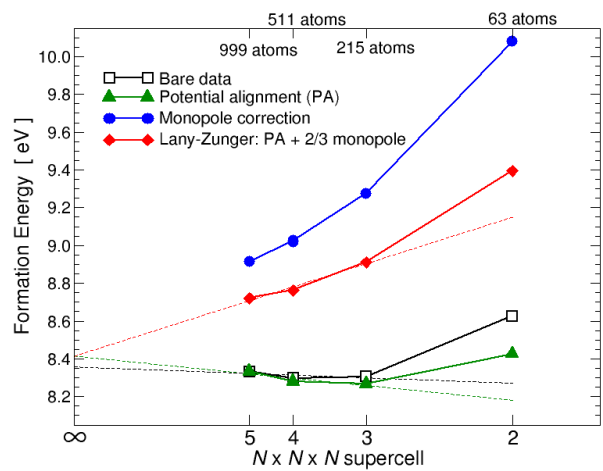}
\caption{(Color online)
Formation energy (silicon-rich) of the silicon vacancy in the charge state 2+ as a function of the supercell size.
We provide the uncorrected data with open square symbols, the potential aligned data with triangle symbols,
the Madelung monopole corrected data with circles,
and the Lany-Zunger corrected data with diamonds.\cite{lany_msmse2009}
In addition, the dashed lines represent reasonable extrapolations based on the three largest supercells.
\label{fig:ef}
}
\end{figure}

The calculations of charged defects in replicated supercells notoriously converge slowly as a function of the supercell size.
The supercell technique gives rise to important finite-size effects, among which the electrostatic interaction between
the charged defects is arguably the largest.
In order to address this issue, a Madelung correction can be designed. \cite{leslie_jpc1985}
However it has been shown in the recent years that a straightforward Madelung correction does more harm than benefit.
\cite{nieminen_msmse2009,lany_msmse2009,freysold_prl2009}
We clarified this issue for our specific system by
performing a careful convergence study about the silicon vacancy \vsi{2+}, which should be the most dramatic case of our study.

In Fig.~\ref{fig:ef}, the formation energy of \vsi{2+} is provided for different cubic supercells ranging from 63 atoms to 999 atoms.
The structure is frozen with all the atoms in the perfect crystal positions except for the four nearest neighbors to the vacancy, whose positions were relaxed in a 63 atoms supercell. This explains why the formation energy is
slightly larger than the fully relaxed energy reported elsewhere in the text.
In the present case, the Madelung correction clearly overestimate the correction.
In Fig.~\ref{fig:ef}, we provide the curve corrected by the popular Lany-Zunger correction scheme \cite{lany_msmse2009}, which consists in a potential alignement
together with a reduced Madelung correction (precisely 2/3 of the usual Madelung term). The Lany-Zunger scheme is also not satisfactory as shown by the figure.
Note that the dielectric constant has been calculated \textit{ab initio} consistently using the same parameters as the defect calculations.
We then found more reliable to perform a mere potential alignment.
Using a 215 atoms supercell with no charge correction allows us to evaluate within 0.1~eV the formation energy of \vsi{2+}.
The values compared to experiments are always differences of formation energies, which are expected to be even more accurate,
since the errors compensate to some extent.

\section{Formation energy of the metastable silicon vacancy and the stable complex}
\label{sec:formation}

\begin{figure}[t]
\includegraphics[width=0.95\columnwidth]{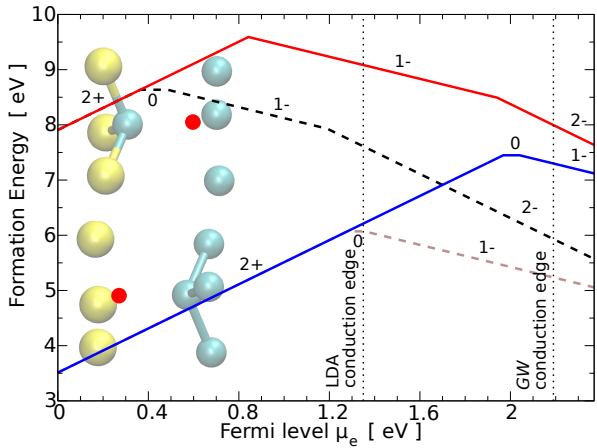}
\caption{(Color online) Formation energy of the silicon vacancy \vsi{} (top)
  and of the complex \vccsi{} (bottom) in 3C-SiC under silicon rich conditions within LDA (dashed lines)
and within the $GW$ approximation (solid lines) as a function of the Fermi level $\mu_e$.
The silicon atoms are light gray (yellow) and the carbon atoms are dark gray (blue).
\label{fig:ef_vsi}
}
\end{figure}

The silicon vacancy is a high energy defect that can only be observed in
heavily irradiated SiC. 
The numerous silicon vacancies created by irradiation are a metastable configuration:
the silicon vacancy experiences a large energy drop when transforming into the complex made of a carbon vacancy and a carbon antisite \vccsi{} \cite{bockstedte_prb2003}.
Furthermore, the experimentally observed silicon vacancies possess a peculiar high-spin configuration (S=3/2) with three aligned spins
resulting in a 1- charge state. This particular spin configuration gives rise to an unambiguous EPR signal, named T1 center.
The T1 center can be observed in irradiated samples, whatever the initial doping conditions
and irradiating particle~\cite{itoh_jap1989,itoh_jem1992}.
After 
irradiation, due to the quantity and the variety of point defects introduced in the material, the donor and acceptor defects
give rise to a compensated system and the Fermi level remains pinned in the vicinity of mid-gap.

In Fig.~\ref{fig:ef_vsi} we provide the formation energy of the two competing configurations for the silicon vacancy:
the original silicon vacancy \vsi{} and the complex \vccsi{}.
The LDA clearly suffers from the band gap problem so that all the transition energies are constrained to be below the LDA conduction edge,
symbolized by a vertical dotted line.
In addition, it would be doubtful to determine the mid-gap region:
would it be the middle of the Kohn-Sham band gap or the middle of the experimental band gap?
Conversely, the $GW$ results offer a much clearer view on the system.
First, the $GW$ transition energies are correctly placed within the full range of the $GW$ band gap,
which matches reasonably well the experimental band gap.
Second, the silicon vacancy adopts a correct 1- charge state in a high-spin configuration for a wide region around mid-gap.
The silicon vacancy appears as metastable with respect to the complex \vccsi{} for any Fermi level in the band gap.
The energy differences are sizable, ranging from 4.4~eV for the p-type region to 0.7~eV for the n-type region.
The $GW$ results show that the complex carries a 2+ charge state in the major part of the band gap.

\section{Proposal for an interpretation of the {\sl E}-line of photoluminescence}
\label{sec:pl}

\begin{figure}[t]
\includegraphics[width=0.95\columnwidth]{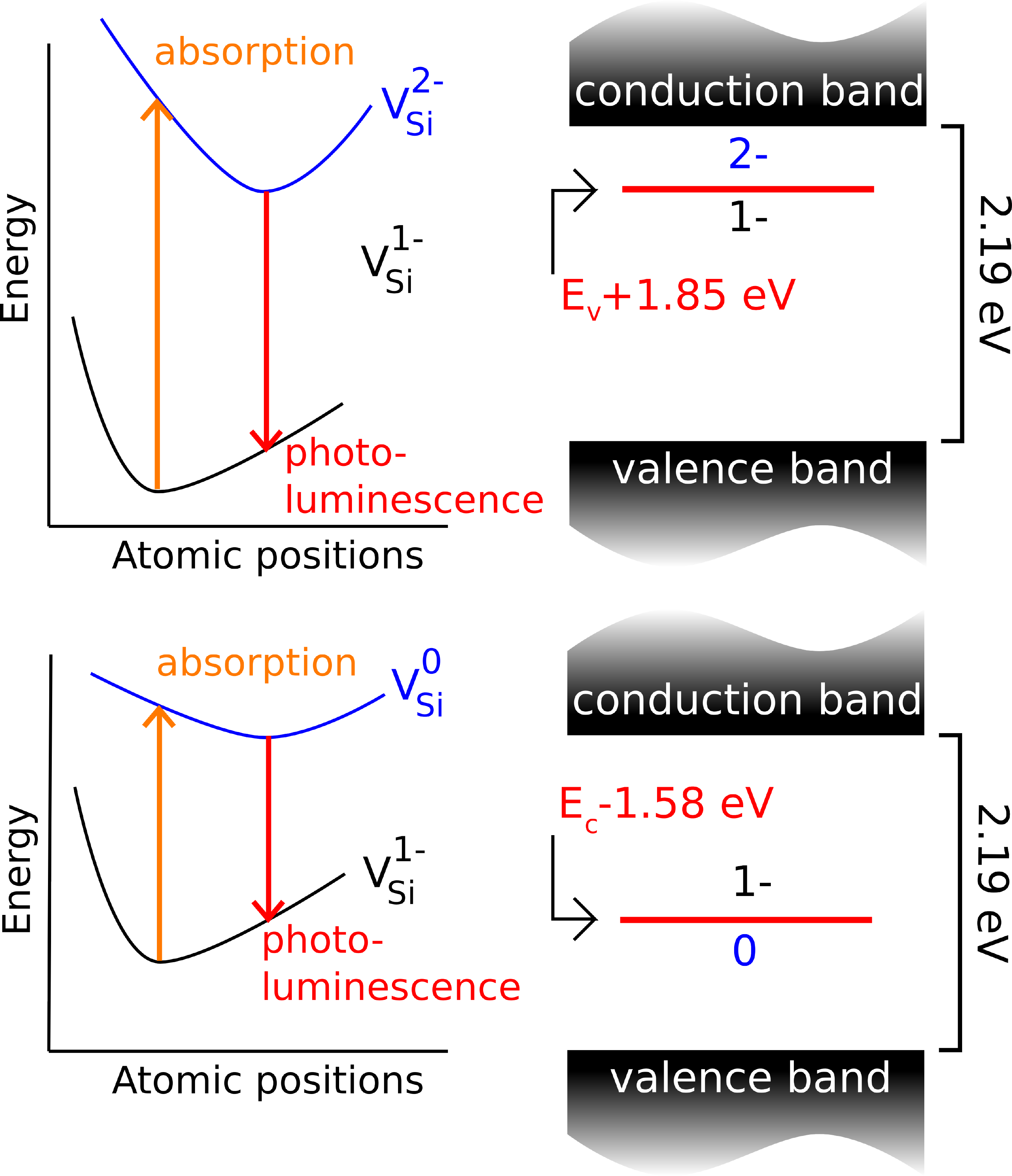}
\caption{(Color online) Transitions associated to the photoluminescence of the silicon vacancy in 3C-SiC.
The schematic Born-Oppenheimer surfaces are represented on the left-hand panels.
The PL levels calculated within the $GW$ approximation are represented on the right-hand panels.
\label{fig:levels}
}
\end{figure}

The zero-phonon line of PL spectroscopy measures the vertical transition energy down from an excited defect,
as schematically shown in the left panels of Fig.~\ref{fig:levels}.
In irradiated samples, additional features appear that can be linked to the intrinsic defects.
In particular, a line, labeled E by Itoh and coworkers \cite{itoh_jap1995}, was convincingly attributed to the silicon vacancy.
Indeed the E line and the above mentioned T1 center in EPR have precisely the same behavior upon thermal annealing: the same three annealing stages.
This clearly demonstrates that the photons of the E line with an energy of 1.91~eV arise from electronic transitions occurring at the silicon vacancy.

In Ref.~\onlinecite{itoh_jap1995}, the authors proposed that the E line comes from the recombination of a conduction electron
with a hole on a silicon vacancy, i.e., \vsi{0}.
So that they tentatively assigned it to the transition level $\epsilon_v(0/1-)$ at an energy of $E_c-1.91$~eV.
This assignment was mainly supported by calculations: for instance, our LDA results place the transition $\epsilon_v(0/1-) = E_v+0.54$~eV.
By combining the LDA transition level together with the \textit{experimental} band gap, 2.37~eV, the resulting emitted photon matches
the experimentally measured photon.

However, this assignment is questioned by our calculations.
When performing higher accuracy $GW$ calculations with the fair $GW$ band gap value,
the obtained vertical transition lies at $\epsilon_v(0/1-) = E_v + 0.79~\text{eV} = E_c-1.58~\text{eV}$, as represented
in Fig.~\ref{fig:levels}.
In this improved framework, the obtained transition energy does not fit the experimental interpretation anymore.
The disagreement is definitely larger than the calculation uncertainties, like the slight underestimation of the band gap by the $GW$ approximation
(2.19~eV instead of 2.37~eV) and the neglect of the excitonic effects.

From our calculations,
another interesting scenario can however be drawn that challenges the previous interpretation of the E line.
In fact, the PL technique is not capable of distinguishing donor-valence recombinations from acceptor-conduction ones.
So that the observed E line may rather arise from the recombination of an
extra electron on the silicon vacancy (i.e., \vsi{2-})
with a hole in the valence bands.
Then, the electronic transition that emits the PL photon will be a signature for the vertical transition $\epsilon_v(2-/1-)$.
Previous evaluations for this transition were biased by the band gap problem: within LDA, $\epsilon_v(2-/1-)=E_v+1.15~\text{eV}$
close to the LDA conduction edge (1.35~eV) and therefore was disregarded for the E line.
Within $GW$, this transition is located interestingly close to the experimental E line, $\epsilon_v(2-/1-)=E_v+1.85~\text{eV}$.
A small error of the photon energy is in fact expected, since our single electron picture does not take into account
the exciton binding energy and since our calculated band gap is slightly too small.
Note that the two sources of error have different signs.
Indeed, the electron-hole attraction always lowers the emitted photon energy.
Exciton binding energies comprised between 0.18 and 0.24~eV were recently
calculated for the carbon vacancy in 4H-SiC~\cite{bockstedte_prl2010}. 
It is likely that an increase of the band gap would push the defect level up, since it is close the conduction edge.
Our data clearly advocate for a reinterpretation of the nature of the E line:
the E line is a signature of the transition from charge state 2- to charge state 1-, with an experimental value of
$\epsilon_v(2-/1-)=E_v+1.91~\text{eV}$ against a $GW$ calculated value of $\epsilon_v(2-/1-)=E_v+1.85~\text{eV}$.

\section{Annealing of the silicon vacancies through a minority charge state}
\label{sec:annealing}

\begin{figure}[t]
\includegraphics[width=0.95\columnwidth]{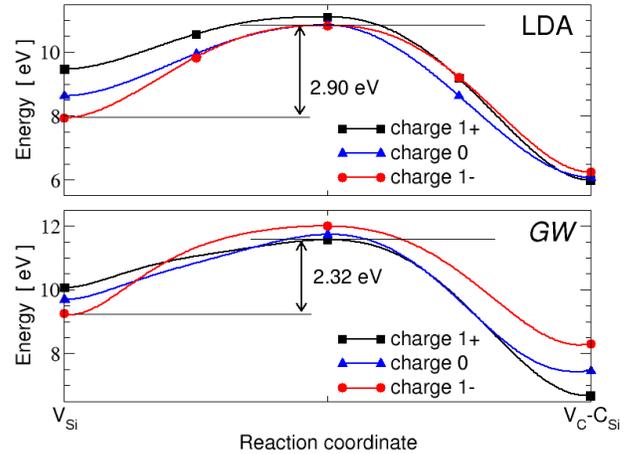}
\caption{(Color online) Energy path of the transformation of \vsi{} into \vccsi{} for different charge states.
The carbon-rich conditions were set.
The Fermi level has been set to the experimental mid-gap value.
The upper panel shows the 5 image CI-NEB path obtained for the 215 atoms supercell within LDA.
The lower panel provides the $GW$ evaluation for the energy of the 3 critical points of the path.
The points actually calculated are depicted with the symbols and the lines are
guides to the eye.
\label{fig:path}
}
\end{figure}


Along with the low temperature EPR and PL studies, the thermal annealing of the silicon vacancy signatures
have been widely studied \cite{itoh_jap1989,itoh_jap1995,lefevre_jap2009}.
The annealing of the silicon vacancy related signals occurs in three stages.
The recovery stages I and II at rather low temperature (400-700~K) anneal
about half of the signal, then stage III at high temperature (1050~K) marks the complete annihilation
of \vsi{}. The activation energy of the last stage was determined to be $E_A=2.2 \pm 0.3$~eV
\cite{itoh_jap1989,lefevre_jap2009}.
The thermal annealing curve yields accurate effective activation energies,
but do not give any clue about the mechanisms at the atomic scale.
In the following we propose a 
picture of the recovery stages of \vsi{}
thanks to the published \textit{ab initio} data and our $GW$ results for a reaction path.

It is reasonable to assume that the first two recovery stages are associated
to the elimination of silicon vacancies by recombination with carbon and
silicon interstitials, whose migration and recombination barriers are in the
range 0.5-1.5~eV~\cite{Bockstedte:2004a,RomaCroc_JNM2010}. We have to exclude from
this process the 4+ charged silicon interstitials, whose migration energy is
expected to be large. Vacancies themselves can also be considered
immobile~\cite{bockstedte_prb2003}. 
The extinction of the T5 center in EPR, assigned to \csplit{}
\cite{petrenko_jpcm2002}, at the same temperature as the first stage suggests
that this is related to the \vsi{} $+$ \csplit{} $\rightarrow
\text{C}_\text{Si}$ recombination. The antisite is a very low energy defect that is very stable once formed.


The third recovery stage shows a first order kinetics according to the isothermal annealing of Ref.~\onlinecite{itoh_jap1989}.
This stage was tentatively proposed as a transformation of the metastable \vsi{} into the lower energy complex \vccsi{}~\cite{Bockstedte:2004a},
but our calculated barriers do not agree with the experimental activation energy as shown in the following discussion.
Again we argue that the calculations suffer from the band gap problem and that the mechanism involved is more complex than was thought previously.

We therefore performed reaction path calculations within LDA for the relevant charge states.
The LDA configurations for the saddle points were then used to build up the $GW$ formation energies
as described above.
The LDA paths were obtained thanks to the climbing image nudged elastic band method (CI-NEB) \cite{henkelman_jcp2000} with 5 images for 215 atoms supercells
as displayed in the upper panel of Fig.~\ref{fig:path}.
In order to compare the different charge states, we set the Fermi level to mid-gap.
One can safely assume that the Fermi level is pinned by the numerous deep donors and
acceptors, which somehow compensate each other in heavily irradiated materials.
In focusing on the Fermi level we implicitly assume that the thermodynamic equilibrium between charge states is always satisfied.
This hypothesis is not true in general. However, in the present case, the annealing experiment is carried out at very high temperature (1050~K)
and the kinetics of charge equilibration is a matter of nanoseconds. We will discuss this point in further details later on.
The obtained LDA transition paths have a minimal barrier of 2.9~eV for the 1- charge state.
This value deviates much from the experimental activation energy.

In the lower panel of Fig.~\ref{fig:path} we provide the $GW$ energies for saddle points and for stable points.
The $GW$ barriers shine light onto a very peculiar diffusion mechanism.
The vast majority of the silicon vacancies possesses a 1- charge state, however the direct transformation into the complex \vccsi{-} is blocked
by a large barrier of 2.75~eV.
The very few positively charged silicon vacancies \vsi{+} experience a low energy barrier of 1.50~eV.
However the total effective activation energy of the reaction in the 1+ charge state is the sum of the barrier energy for charge 1+
and of the energy needed to turn the vacancy from 1- to 1+.
We finally obtain an activation energy of 2.32~eV, which lies within the
experimental uncertainty. Some lowering of this figure by temperature effects
can be further expected.
The transformation of the metastable silicon vacancy into the stable complex is mediated by a double charge change.
This extinction of \vsi{} is interestingly driven by an ultra minor charge state:
even at 1050~K, the concentration of the 1+ charge state is only [\vsi{+}]/[\vsi{-}] $\sim 10^{-4}$.
Note that only the $GW$ values for the formation energies of \vsi{} and of the barrier heights were able to explain the experimental observations.

Let us now assess the validity of instantaneous equilibrium between the different charge states of \vsi{}.
All our calculated paths are based on this assumption.
Indeed, this hypothesis can be validated by experimental data from Deep Level Transient Spectroscopy (DLTS) on a
very similar system, hexagonal SiC (4H-SiC). 4H-SiC is expected to resemble much 3C-SiC, except for the band gap which is noticeably larger.
DLTS experiments performed in heavily irradiated 4H-SiC measure a gigantic timescale to restore charge equilibrium of about two weeks for deep levels
at room temperature. \cite{lebedev_jap2000}
These levels have been measured 1.2~eV away from the band edges of 4H-SiC, which constitutes an upper bound for a deep level in 3C-SiC,
for which the band gap is only 2.37~eV.

The theory underlying DLTS specifies that the recovery time $\tau$ follows an exponential law: \cite{bourgoin}
\begin{equation}
\label{eq:tau}
 \tau \propto \exp \left( \Delta E / k_B T \right) ,
\end{equation}
where $\Delta E$ is the energy difference between the defect level and the closest band edge, $k_B$ is the Boltzmann constant and $T$ the temperature.
In the annealing experiments of Itoh and coworkers \cite{itoh_jap1989}, the critical temperature is 1050~K.
For this high temperature, the exponential law in Eq.~(\ref{eq:tau}) gives a surprising collapse of the recovery time, which value is only 2~ns.
The annealing experiment usually lasts for several minutes or hours. As a consequence, the charge re-equilibration is many orders of magnitude
faster than the annealing experiment itself. The instantaneous charge equilibration can be safely assumed as we did to draw Fig.~\ref{fig:path}.

\section{Conclusions}

As a conclusion, the use of the state-of-art $GW$ approximation for large supercells of 215 atoms allowed us to completely clarify
the experimental observations concerning the silicon vacancy in 3C-SiC.
The calculated properties deviate noticeably from the previous LDA studies, owing much to the absence of the band gap problem.
Beside the nice agreement with the experimental data, we brought deeper understanding in the physics of the silicon vacancy:
we proposed a new assignment for the main PL line, which should be attributed to the 2-/1- charge transition;
we confirmed that the third recovery stage of \vsi{} is indeed a transformation into \vccsi{}
and proposed a mechanism that channels through the very rare 1+ charge state.
This last application was possible only using $GW$ calculations for the reaction saddle point.

\begin{acknowledgments}
We acknowledge useful discussions with M. Bockstedte, J.-P. Crocombette, and Y. Limoge.
The present calculations were based on the \textsc{abinit} code \cite{abinit_cpc2009} for the $GW$ calculations
and on the Quantum-espresso code \cite{quantum-espresso} for the CI-NEB paths.
This work was performed using HPC resources from GENCI-CINES and GENCI-CCRT (Grant 2010-gen6018).
\end{acknowledgments}


%

\end{document}